
\def\m@th{\mathsurround=0pt}
\newif\ifdtpt
\def\displ@y{\openup1\jot\m@th
    \everycr{\noalign{\ifdtpt\dt@pfalse
    \vskip-\lineskiplimit \vskip\normallineskiplimit
    \else \penalty\interdisplaylinepenalty \fi}}}
\def\eqalignc#1{\,\vcenter{\openup1\jot\m@th
                \ialign{\strut\hfil$\displaystyle{##}$\hfil&
                              \hfil$\displaystyle{{}##}$\hfil&
                              \hfil$\displaystyle{{}##}$\hfil&
                              \hfil$\displaystyle{{}##}$\hfil&
                              \hfil$\displaystyle{{}##}$\hfil\crcr#1\crcr}}\,}
\def\eqalignnoc#1{\displ@y\tabskip\centering \halign to \displaywidth{
                  \hfil$\displaystyle{##}$\hfil\tabskip=0pt &
                  \hfil$\displaystyle{{}##}$\hfil\tabskip=0pt &
                  \hfil$\displaystyle{{}##}$\hfil\tabskip=0pt &
                  \hfil$\displaystyle{{}##}$\hfil\tabskip=0pt &
                  \hfil$\displaystyle{{}##}$\hfil\tabskip\centering &
                  \llap{$##$}\tabskip=0pt \crcr#1\crcr}}
\def\leqalignnoc#1{\displ@y\tabskip\centering \halign to \displaywidth{
                  \hfil$\displaystyle{##}$\hfil\tabskip=0pt &
                  \hfil$\displaystyle{{}##}$\hfil\tabskip=0pt &
                  \hfil$\displaystyle{{}##}$\hfil\tabskip=0pt &
                  \hfil$\displaystyle{{}##}$\hfil\tabskip=0pt &
                  \hfil$\displaystyle{{}##}$\hfil\tabskip\centering &
                  \kern-\displaywidth\rlap{$##$}\tabskip=\displaywidth
                  \crcr#1\crcr}}
%

%


\def\charlvmidlw#1#2{\,\vtop{\ialign{##\crcr
      #1\crcr\noalign{\kern1pt\nointerlineskip}
      $\hfil#2\hfil$\crcr}}\,}
\def\charlvlowlw#1#2{\,\vtop{\ialign{##\crcr
      $\hfil#1\hfil$\crcr\noalign{\kern1pt\nointerlineskip}
      #2\crcr}}\,}
\def\charlvmidup#1#2{\,\vbox{\ialign{##\crcr
      $\hfil#1\hfil$\crcr\noalign{\kern1pt\nointerlineskip}
      #2\crcr}}\,}
\def\charlvupup#1#2{\,\vbox{\ialign{##\crcr
      #1\crcr\noalign{\kern1pt\nointerlineskip}
      $\hfil#2\hfil$\crcr}}\,}

\def\vspce{\kern4pt} \def\hspce{\kern4pt}    

\def\emptybox{\vbox{\kern.7ex\hbox{\kern.5em}\kern.7ex}}
 \font\sevmi  = cmmi7              
    \skewchar\sevmi ='177
 \font\fivmi  = cmmi5              
    \skewchar\fivmi ='177
\font\tenmib=cmmib10
\newfam\bfmitfam

\textfont\bfmitfam=\tenmib
\scriptfont\bfmitfam=\sevmi
\scriptscriptfont\bfmitfam=\fivmi
%

%
\def\mathcedilla{\vtop{\hbox{c}{\kern0pt\nointerlineskip}
	         {\hbox{$\mkern-2mu \mathchar"0018\mkern-2mu$}}}}

\mathchardef\gq="0060
\mathchardef\dq="0027
\mathchardef\ssmath="19
\mathchardef\aemath="1A
\mathchardef\oemath="1B
\mathchardef\omath="1C
\mathchardef\AEmath="1D
\mathchardef\OEmath="1E
\mathchardef\Omath="1F
\mathchardef\imath="10
\mathchardef\fmath="0166
\mathchardef\gmath="0167
\mathchardef\vmath="0176

%

\def\twodot{.\kern-0.1em.}

\def\paral{\mathrel{/\kern-.25em/}}
\def\grlo{\mathrel{\hbox{\lower.2ex\hbox{\rlap{$>$}\raise1ex\hbox{$<$}}}}}
\def\logr{\mathrel{\hbox{\lower.2ex\hbox{\rlap{$<$}\raise1ex\hbox{$>$}}}}}
\def\greq{\mathrel{\hbox{\lower1ex\hbox{\rlap{$=$}\raise1.2ex\hbox{$>$}}}}}
\def\loeq{\mathrel{\hbox{\lower1ex\hbox{\rlap{$=$}\raise1.2ex\hbox{$<$}}}}}
\def\grsim{\mathrel{\hbox{\lower1ex\hbox{\rlap{$\sim$}\raise1ex\hbox{$>$}}}}}
\def\losim{\mathrel{\hbox{\lower1ex\hbox{\rlap{$\sim$}\raise1ex\hbox{$<$}}}}}
%
\font\ninerm=cmr9
\def\uniset{\rlap{\ninerm 1}\kern.15em 1}

\def\emptysq{\mathbin{\vbox{\hrule\hbox{\vrule height1ex \kern.5em
                            \vrule height1ex}\hrule}}}
\def\emptyrect{\mathbin{\vbox{\hrule\hbox{\vrule height1ex \kern1em
                              \vrule height1ex}\hrule}}}
\def\rightonleftarrow{\mathrel{\hbox{\raise.5ex\hbox{$\rightarrow$}\ignorespaces
                                   \lower.5ex\hbox{\llap{$\leftarrow$}}}}}
\def\leftonrightarrow{\mathrel{\hbox{\raise.5ex\hbox{$\leftarrow$}\ignorespaces
                                   \lower.5ex\hbox{\llap{$\rightarrow$}}}}}

\def\bkB{{\rm I\kern-.17em B}}
\def\bkC{{\rm \kern.24em
            \vrule width.05em height1.4ex depth-.05ex
            \kern-.26em C}}
\def\bkD{{\rm I\kern-.17em D}}
\def\bkE{{\rm I\kern-.17em E}}
\def\bkF{{\rm I\kern-.17em F}}
\def\bkG{{\rm \kern.24em
            \vrule width.05em height1.4ex depth-.05ex
            \kern-.26em G}}
\def\bkH{{\rm I\kern-.22em H}}
\def\bkI{{\rm I\kern-.22em I}}
\def\bkJ{{\rm \kern.19em
            \vrule width.02em height1.5ex depth0ex
            \kern-.20em J}}
\def\bkK{{\rm I\kern-.22em K}}
\def\bkL{{\rm I\kern-.17em L}}
\def\bkM{{\rm I\kern-.22em M}}
\def\bkN{{\rm I\kern-.20em N}}
\def\bkO{{\rm \kern.24em
            \vrule width.05em height1.4ex depth-.05ex
            \kern-.26em O}}
\def\bkP{{\rm I\kern-.17em P}}
\def\bkQ{{\rm \kern.24em
            \vrule width.05em height1.4ex depth-.05ex
            \kern-.26em Q}}
\def\bkR{{\rm I\kern-.17em R}}
\def\bkT{{\rm \kern.24em
            \vrule width.02em height1.5ex depth 0ex
            \kern-.27em T}}
\def\bkU{{\rm \kern.30em
            \vrule width.02em height1.47ex depth-.05ex
            \kern-.32em U}}
\def\bkZ{{\rm Z\kern-.32em Z}}
%


\magnification=1200
\baselineskip=17pt
\overfullrule=0pt
\vskip 25pt
\centerline{{\bf SUPERSYMMETRY BREAKING IN THE NAMBU-JONA-LASINIO APPROACH}}
\vskip 12pt
\centerline{by}
\vskip 12pt
\centerline{R. Peschanski and C.A. Savoy}
\centerline{{\sl CEA/DSM/Service de Physique Th\'eorique\/}}
\centerline{{\sl CE-Saclay, F-91191 Gif-sur-Yvette Cedex, FRANCE\/}}
\vskip 17pt
\centerline{{\bf ABSTRACT}}
\vskip 17pt
Gaugino condensation in the hidden sector of supergravity models is
described within a Nambu-Jona-Lasinio type of approach by minimization of a
one-loop scalar potential. The essential ingredients of the mechanism are
auxiliary superfields whose v.e.v. generate gaugino
condensation and supersymmetry breaking, introduced through
Lagrange multipliers.
For phenomenologically acceptable values of the gauge couplings, gaugino
condensation is disfavoured in this approach. For completeness, it is shown
that supersymmetry breaking would occur for a stronger coupling, but at a
scale inconsistent with the expectations.
\vfill\eject

 In supergravity models, the spontaneous breaking of local supersymmetry,
or super-Higgs mechanism may generate
soft supersymmetry breaking  terms that allow to fulfill such phenomenological
requirements as scalar masses and spontaneous gauge symmetry breaking at
scales of $ O(\rm{TeV}) $.
However, the super Higgs mechanism implies the existence of a supergravity
breaking scale, intermediate between $ M_{ {\rm Planck}} $ (or $ M_c, $ the
grand-unification
scale) and $ M_{Z^0}. $ An intermediate scale expected to be of $
O(10^{13} {\rm GeV} ) $ seems
difficult to implement in a natural way, except for the mechanism of
{\sl gaugino
condensation\/}$ ^{[1]} $ in a \lq\lq hidden\rq\rq\ sector gravitationally
coupled to the
\lq\lq observable\rq\rq\ sector of physical matter and gauge fields.
Gaugino condensation is expected to occur from the renormalization group flow
of the coupling constant from the unification scale downwards. Indeed,
with an appropriate large gauge group in the hidden sector
one expects condensates to form at the
suitable scale.
One
interest of this
mechanism is its compatibility$ ^{[2]} $ with supergravity models coming from
superstring unification, where the absence of tree-level intermediate scales
and the existence of large hidden gauge sectors follows  from the
heteretotic string constructions.

Gaugino condensation  has been
derived for globally supersymmetric theories$ ^{[3]}$ within both the effective
potential and the instanton approaches. Thus, a global chiral
symmetry is found to be spontaneously broken.  However, it
does not lead to supersymmetry breaking unless one
includes   a dilaton-axion supermultiplet coupled to the
gauge sector, the so called $ S $-fields$ ^{[3,4]}. $ Interestingly enough,
this
pattern is indeed suggested in string theory.

Basically, only effective Lagrangian approaches have been extended
to the supergravity models, in order to justify an $ S $-dependent
effective superpotential leading to the super Higgs mechanism.
Successive attempts to formulate gaugino condensation in supergravity
have been confronted with difficulties related to the non-perturbative
aspects of the condensation phenomenon. Thus, the minimization of
the effective potential leads to the $ S \to \infty $ limit$ ^{[3,4]}.$

The original proposal to stabilize the dilaton $v.e.v.$ within the
superstring framework has been challenged because of $ T $-duality
requirements$ ^{[5]} $.
Indeed,
the presence of \lq\lq moduli\rq\rq\ (so-called $ T $-fields) in
superstring compactification is quite
relevant  as they entail the \lq\lq
no-scale\rq\rq\
structure of the theory.
A mechanism based on the existence of two or more condensates
whose interference leads to a minimum with a finite dilaton $v.e.v.$ has been
put
forward$ ^{[6]}. $ However, it is not clear how well the different scales
introduced in this case can be determined. Moreover, it has been argued
that supersymmetry would not be broken
$ ^{[7]} $ at the minimum of the scalar potential.

This discussion is the main motivation for considering the Nambu Jona- Lasinio
approach$
^{[8]} $ to
gaugino condensation$ ^{[9]}.$ Indeed, it has been found in Ref.$[9]$
that, under some assumptions on the effective supergravity lagrangian,
gaugino condensation and supersymmetry breaking are obtained. Furthermore, the
dilaton v.e.v. turns out to be consistent with the phenomenological value
of the gauge coupling.

In this paper we show through a more careful treatment of the
effective lagrangian in the NJL approach, that supersymmetry breaking would
require too large values of the gauge couplings (even in the case of more
than one condensate).

In the
most appropriate version of the NJL approach one introduces an auxiliary
composite field $ \sum^{ }_ a \left(\lambda^ a\lambda^ a \right) $ $ (\lambda^
a $ are the gaugino fields).
The one-loop effective potential is then evaluated and provides the
non-trivial dependence on the composite field. However, this effective
potential is quadratically divergent and defined by the introduction of a
cut-off. Then, the minimization of the one-loop potential gives a non-trivial
vacuum expectation value to the composite field, in the form of a \lq\lq gap
equation\rq\rq , for couplings above some critical value defined in terms of
the
cut-off. In this way, the fermion chiral symmetry is broken and so are other
symmetries acting non-trivially on the fermion (i.e., gaugino) bilinear.
Therefore the one-loop quantum correction is implementing the necessary
dynamical dependence on the composite fields. Moreover, kinetic terms are
also generated at one loop, supporting the physical interpretation that the
composite field become propagating below the critical scale defined by the
cut-off. All this can be made precise in the large-$ N $ limit, where $ N $ is
the
number of fermions (gauginos).

The supersymmetric version of the NJL is somewhat more complex but it has
been widely discussed in the literature$ ^{[10]}, $ e.g., in the framework of
top
condensation$ ^{[11]}. $ In this context, one considers \lq\lq
super-composite\rq\rq\ chiral
fields $ Q^cQ $ and soft supersymmetry breaking terms are added by hand.
Instead,
gaugino condensates are defined as the first components of bilinears, $
\left(W^\alpha_ aW^\alpha_ \alpha \right)_{\theta =0}, $
where $ W_\alpha $ is the spinor chiral field that contain the gauge strength
among its
components. One aims to induce the dynamical breaking of
supersymmetry from non-vanishing last components $ \left(W^\alpha_ aW^a_\alpha
\right)_{\theta^ 2} $ in the wake of the
symmetry breaking by the first components. A further complication is introduced
by the fact that there could be several orders of magnitude between the
unification scale where the theory is settled and the condensation scale under
investigation. Since condensation is presumably produced as a
non-perturbative phenomenon by the increasing of the gauge coupling as the
energy scale decreases, this scale dependence has to be dynamically
accounted for in the
NJL effective theory.

As stated above, the relevant framework in this study is
supergravity, enriched by a dilaton-dilatino-axion sector and some
modulus superfields suggested by superstring theory. The superfields contain
auxiliary fields $ (F $ components) that control supersymmetry breaking.
Therefore we choose to keep these auxiliary components in the calculation of
the one-loop effective potential and eliminate them at the end from the
overall (cut-off dependent) potential.

Let us assume a hidden gauge sector with a semi simple gauge group $ G=
\bigotimes^ n_{i=1}G_i $
and dimension $ N_G= \sum^ n_{i=1}N_i $ where $ N_i $ is the dimension (i.e.,
the number of
gaugino) of each one of the $ n $ factors $ G_i $ in $ G. $ Let us neglect all
matter
fields, all gauge superfields, and any additional component of the observable
sector that are irrelevant for the discussion hereafter. Instead the dilaton
chiral superfield $ S $ play a major r\^ole in supersymmetry breaking and it
must
be retained together with the essential part of the moduli fields. At the
level of our approach it is enough to introduce only one chiral superfield $
T, $
the overall modulus field. Correspondingly, we set the relevant part of the
Kahler potential to the usual tree-level expression:
$$ K(S,T)=- {\rm ln} \left(S+S^{\dag} \right)-3\ {\rm ln} \left(T+T^{\dag}
\right) \eqno
(1) $$
We take $ M_{ {\rm
Planck}}=1 $ as our
unity for energy scales unless contrarily stated.

The supersymmetric
Yang-Mills lagrangian is
$$ \left[{S \over 4} {\rm Tr} \ W^\alpha W_\alpha \right]_{\theta^ 2}+ {\rm
h.c.} \eqno (2) $$
where $ W^\alpha $ are the chiral (Majorana) spinor superfield whose first
components are
the gaugino fields $ \lambda^ \alpha_{ a_i} $ $ (a_i=1,...,N_i; $ $ i=1,...,n)
$ in the hidden sector. The
v.e.v. of the dilaton component of $ S $ sets the gauge coupling at the
compactification scale $ M_c $ as follows:
$$ S+S^\ast =2g^{-2} \left(M_c \right) \eqno (3) $$
$$ M_c={2M_{ {\rm Planck}} \over \left(S+S^\ast \right)^{1/2} \left(T+T^\ast
\right)^{1/2}} \eqno  $$

\noindent For simplicity we are assuming a unique gauge coupling $ g^2
\left(M_c \right)
$ for all the
factors $ G_i. $ This amounts to all the levels of their Kac-Moody
algebra
being equal to 1, but the generalization to $ k_i\not= 1 $ is obvious. However,
those
couplings will be splitted by the RG running. At one-loop they obey the
equations:
$$ \eqalignno{{ 1 \over g^2_i(\Lambda )} & ={1 \over g^2 \left(M_c
\right)}+4\beta_ i\ {\rm ln}{\Lambda^ 3 \over M^3_c} &  \cr \beta_i & ={C_2
\left(G_i \right) \over 32\pi^ 2} & (4) \cr} $$
where $ C_2 \left(G_i \right) $ is the Casimir eigenvalue of $ adj \left(G_i
\right). $

In order to describe gaugino condensation and to determine $ \left\langle
\sum_{a,\alpha} \lambda^\alpha_{ a, i}\lambda_{ \alpha a, i} \right\rangle
\equiv \left\langle
\lambda^a \lambda_a \right\rangle_ i $
we introduce\footnote{$
^1 $
}{Notice however that $ {\rm Tr}_iW^\alpha W_\alpha $ is a constrained chiral
superfield. This fact
has motivated a different approach to this problem based on the linear
supermultiplet formalism$ ^{[12]}. $} the auxiliary chiral superfields $ U_i $
and the conjugate chiral superfields $ X_i $ and we add to the
lagrangian the following expression
$$ \sum^ n_{i=1}X_i \left(U_i- {\rm Tr}_iW^\alpha W_\alpha \right) \eqno (5)
$$
The functional integration over the $X_i$' s that act as Lagrange multipliers
followed by the integration over the superfields $U_i$' s gives back the
original action.
The v.e.v.'s $ \left\langle U_i \right\rangle $ set the condensation scale for
each sector $ G_i $ of the hidden
gauge sector.

On quite general grounds, the NJL approach to spontaneous symmetry breaking
amounts to a minimization of the effective scalar potential for the auxiliary
scalars associated to the fermion condensates. The effective potential
calculated at one-loop depends on a cut-off scale $ \Lambda . $
The
couplings entering this effective potential should be calculated at the
condensation scale of
$ {\cal O}\left( \left\langle U \right\rangle \right). $ The gauge couplings
explicitly appear in the masses of gaugino $ M_i(\langle U\rangle) \propto
g^2_i, $
 that circulate around the loop. Equation (4) suggests to replace  (2) by
$$ \left\{ \left[{S \over 4}+\beta_ i\ {\rm ln}{U_i \over M^3_c} \right] {\rm
Tr}_iW^\alpha W_\alpha \right\}_{ \theta^ 2}+ {\rm h.c.} \eqno (2^{\prime} )
$$
Consistently, if (5) is used to formally eliminate the $W^\alpha$ superfields
the logarithm term in (2') reproduces their contribution to the superconformal
anomaly$ ^{[3,4]}.$

The supergravity theory defined by the lagrangian given by (1), $ (2^{\prime}
) $ and (5)
can be now contemplated in two forms, that are equivalent as a  matter of
fact:

a)\nobreak\ From $ (2^{\prime} ) $ and (5) one derives the superpotential
$$ W= \sum^{ }_ iX_iU_i \eqno (6) $$
and the following expression for the field dependent  gaugino masses
$$ \eqalignno{ M_i & ={ \left(\hat X_i \right)_{\theta^ 2} \over\hat X_i+\hat
X^\ast_ i}={\hat F^i_X \over\hat X_i+\hat X^\ast_ i} &  \cr\hat X_i & \equiv{
S \over 4}+X_i+\beta_ i\ {\rm ln}{U_i \over M^3_c} & (7) \cr} $$

b)\nobreak\ Because (5) defines $ n $ $\delta-$functions, one can replace
$ {\rm Tr}_iW^\alpha W_\alpha $ by $ U_i $
in $ (2^{\prime} ). $ In this form, that turns out to be handier than a), the
superpotential becomes:
$$ W= \sum^{ }_ iU_i \left({S \over 4}+X_i+\beta_ i\ {\rm ln}{U_i \over M^3_c}
\right) \eqno (8) $$
and the gaugino masses reduce to
$$ M_i={F^i_X \over X_i+X^\ast_ i} \eqno (9) $$
where $ F^i_X $ are the auxiliary fields in the chiral multiplets $ X_i. $

Let us formulate the NJL approach in the more convenient form b). The
results in a) can be checked to be  identical with the replacements $ F^i_X
\longrightarrow\hat F^i_X, $ $ X_i \longrightarrow\hat X_i, $ as insured
by the implicit $\delta-$functions implied by the contributions (5) to
the lagrangian.

The scalar potential at one-loop, $ V=V_0+V_1, $ is obtained from (1), (8) and
(9)
as follows:

(i)\nobreak\ $ V_0 $ is given by the classical supergravity lagrangian$ ^{[13
]}: $
$$ V_0=-3\left\vert m_{3/2} \right\vert^ 2-
{ \left\vert F^S \right\vert^
2 \over \left\vert S+S^\ast\right\vert^2}
-{ 3 \left\vert F^T \right\vert^
2 \over \left\vert T+T^\ast\right\vert^3}
- \left[m_{3/2} \sum^{ }_{ I=S,T,U_i,X_i}F^I{\partial \over
\partial I}(K+ {\rm ln}\ \eta^{-6}(T)\ W)+ {\rm h.c.} \right] \eqno (10) $$
where $ m_{3/2} $ is the gravitino mass,
$$ m_{3/2}={\eta^{-6}(T)\ W \over \left(S+S^\ast \right)^{1/2} \left(T+T^\ast
\right)^{3/2}}
\eqno (11) $$

The introduction of the Dedekind function $\eta (T)$ ensures the modular
invariance of the lagrangian$ ^{[14,5]}.$ In this way the no-scale vaccuum
degeneracy is lifted and the $ T$ v.e.v. is fixed. Thus, the duality-invariant
 value $ T = 1$ is always an extremum of the scalar potential but not
necessarily the absolute minimum.

(ii)\nobreak\ The one-loop contribution $ V_1 $ to the effective potential is
obtained by
the addition of the contribution of all the  physical particles$ ^{[15]}: $
$$ V_1=-2 \sum^{ }_ iN_i\ H \left( \left\vert M_i \right\vert^ 2,\Lambda
\right)- \sum^{ }_ A\nu_ AH \left(\kappa_A \left\vert m_{3/2} \right\vert^
2,\Lambda \right) \eqno (12) $$
where $ H $ is defined by the cut-off integral
$$ H \left(m^2,\Lambda \right)= \int^\Lambda{ {\rm d}^4p \over (2\pi )^4}
{\rm ln} \left(p^2+m^2 \right)
  ={\Lambda^ 4 \over 64\pi^ 2} \left[{m^2 \over \Lambda^ 2}+ {\rm
ln} \left(1+{m^2 \over \Lambda^ 2} \right)-{m^4 \over \Lambda^ 4} {\rm ln}
\left(1+{\Lambda^ 2 \over m^2} \right) \right]
 \eqno (13) $$
and $ m $ is the mass of the state as a function of the scalar fields. The
cut-off $ \Lambda $ defines the physical region for the composite auxiliary
fields, and
the sums in (12) include all propagating fields such that $ m<\Lambda . $ The
gaugino
masses $ M_i $ are given by (9). In the second term $ \nu_ A $ are
multiplicities (and signs) from
states whose mass is $ \sqrt{ \kappa_A} \left\vert m_{3/2} \right\vert . $ The
factors $ \kappa_A $ are assumed to be independent
of the fields $ X_i,U_i. $ Thus, the gravitino sector contributes to $ V_1 $
with the
well-known$ ^{[16]} $ term,
$$ 2H \left(4\left\vert m^2_{3/2}\right\vert \right)-4H \left(\left\vert
m^2_{3/2}\right\vert \right) \eqno (14) $$
The NJL is better justified in the large-$ N $ limit. For this reason, the
gaugino contribution in the one-loop potential is instrumental while other
particle contributions seem circumstantial. However we shall keep (12) as it
is for completeness and for some further discussion.

The essence of the present approach is the non-trivial dependence of $ V_1 $
on
the auxiliary fields $ F^i_X $ $ (\hat F^i_X $ in form a)) in the gaugino
masses.
Thus, these Lagrange multipliers acquire a dynamical r\^ole and contribute to
the vacuum energy. This is the effective-potential reflexion of the NJL
transmutation of the auxiliary fields into dynamical composite fields below
some critical scale characterized by the cut-off $ \Lambda . $

The next step is the minimization of $ V=V_0+V_1 $ with respect to the fields.
Let us start with the elimination of the auxiliary fields, $F_I \ (I =
S,T,U^i,X^i).$

(i)\nobreak\ $ \partial V/\partial F^i_U=0, $ express $n$ constraints
that relate the condensate scales
to the unification mass, through
the equations,
$$ {\partial W \over \partial U_i}=0\ ,\ \ \ \ \ \ \ \ \ \ (i=1,...,n) \eqno
(15) $$
which gives with the superpotential (8),
$$ U_i=M^3_c\ {\rm exp} \left(-{S \over 4\beta_ i}-{X_i \over  \beta_ i}-1
\right) \eqno (16) $$
as well as the following expression for the superpotential at the extrema of
the scalar potential,
$$ W=- \sum^{ }_ i\beta_ iU_i \eqno (17) $$
Notice that (15) and (11) entail
$$ {\partial m_{3/2} \over \partial U_i}=0 \eqno (18) $$
These zeroes in the Jacobian matrix prevent us from using $ m_{3/2} $ as a
minimization variable if supersymmetry breaking follows the present pattern.
Since this is an ordinary implement in studies of supersymmetric theories,
this warning is noticeworthy.

(ii)\nobreak\ $ \partial V/\partial F_S=0, $ eliminates the auxiliary
component of the dilaton superfield
which using (17) can be written as
$$ {F_S^\ast \over S+S^\ast} =m_{3/2} \left[1+{ \left(S+S^\ast \right) \sum^{
}_
iU_ i \over 4 \sum^{ }_ i\beta_ iU^\ast_ i} \right] \eqno (19) $$

(iii)\nobreak\ $ \partial V/\partial F_T=0, $ eliminates the auxiliary
component of the modulus (moduli, in the more general case):
$$ {F_T^\ast \over (T+T^\ast)^2} =m_{3/2} \left({1 \over
T+T^\ast } + 2\ {\eta'(T) \over \eta (T)}
 \right) \eqno (20) $$

(iv)\nobreak\ $ \partial V/\partial F^i_X=0 $ yield  the \lq\lq gap
equations\rq\rq corresponding to a saddle-point integration,
$$ {M_i U_i \left(X_i+X^\ast_ i \right) \over \left(S+S^\ast \right)^{1/2}
\left(T+T^\ast \right)^{1/2}}+2N_i\left\vert M^2_ i\right\vert H^{\prime}
\left( \left\vert M_i^2
\right\vert,\Lambda \right)=0\ ,\ \ \ \ \ \ \ \ \ \ (i=1,...,n) \eqno (21) $$
$$ H^{\prime} \left(m^2,\Lambda \right)={ {\rm d} \over {\rm d} m^2}H
\left(m^2,\Lambda \right) \eqno  $$
This equation is the core of the NJL approach. Indeed it corresponds
to the minimization of the one-loop vaccuum energy with respect to
 the
gaugino mass (9) considered as a dynamical variable\footnote{$ ^2 $}{
A manifest problem in the above relations concerns the dependence on $ M_i, $
thus
on the $ F^i_X $ auxiliary fields.
The usual couplings of chiral multiplets to supergravity only produce
lagrangians quadratic in $ F $'s and in the scalar field derivatives.
 Instead, $ F^i_X $'s appear in the potential
(12)
within the cut-off integral (13).
This dependence could arise the question of the supersymmetric consistency of
the NJL approach.
At one-loop level the two options for eliminating the $ F-$
components, namely before or after the calculation of the one-loop theory
should produce identical results to lowest order in $ \hbar .$ The
question should be settled by a complete calculation of the
one-loop action.
We
thank M.K. Gaillard for a discussion on
this
problem.}
. In this respect
our treatment is quite different\footnote{$ ^3 $}{
In the absence of  Lagrange multipliers, the form b) corresponding
to (8) cannot be introduced so that for a comparison with Ref.[9]
our equivalent form a) is more suitable.}
from that in Ref.[9] where the Lagrange
superfields $X^i$ is replaced by a (first component) constant which
is estimated by physical arguments. In summary, the main difference
with Ref.[9] is that in our case
the equivalence between $ U $ and $\left( W^\alpha W_\alpha \right)$ is
imposed as a functional constraint while in Ref.[9]
it is only a on-shell relation. This
has important impact on the results as we now turn to discuss.

After the integration on the auxiliary fields, the scalar potential becomes:
$$\eqalignno{
\hat V &= \left\vert m_{3/2}\right\vert^2
 \left[ -3 + 3 \left\vert g(T)\right\vert^2 + \left\vert \omega
\right\vert^2\right] &  \cr
&\ \ \ \       \ + 2 \sum_i^{ } N_i \left[ 2 \left\vert M_i\right\vert^2
H'\left(\left\vert
M_i\right\vert^2\right)
 - H\left(\left\vert M_i\right\vert^2\right)\right]
 -\sum_A^{ } \nu_A H\left(\kappa_A\left\vert m_{3/2}^2\right\vert\right)
 &  \cr
&\ \ \ \ \ \ \ \ \ \left\vert g(T)\right\vert^2  = \left\vert 1+ 2\ (T+T^\ast)
{\eta'(T)
\over \eta (T)}\right\vert^2  & \cr
&\ \ \ \ \ \ \ \ \ \left\vert\omega\right\vert^2  = \left\vert 1 - {S+S^\ast
\over 4 W}
\sum_i^{ } U_i\right\vert^2  & (22)
\cr}
$$
It is to be minimized with respect to the variables
$T,U^i,X^i$ with the constraints (15), which fix the variables $U^i$ to be of
the form (16).
For simplicity, we retain only the dominant terms in the functions $ H $ and $
H^{\prime} $ in
(12)-(21)
$$ \eqalignno{ H(m^2,\Lambda ) & \sim \rho^ 2m^2 &  \cr H^{\prime} (m^2,\Lambda
) &
\sim \rho^ 2\ \ \ \ \ \ \ \ \rho^ 2\equiv{ \Lambda^ 2 \over 32\pi^ 2} &  \cr
&  (m^2\ll \Lambda^ 2) & (23) \cr} $$
Our conclusions will be independent of this approximation as far as
$M_i < \Lambda,$ which is a necessary condition for the whole approach.
Indeed, the sign of the gaugino contribution to the scalar potential (22)
only flips at $M_i > 8\Lambda.$

Let us begin with the modulus $T.$ The symmetric point $T=1$ corresponds to the
minimum of the potential if ( in the approximation (23) for $H$)
$$
\left\vert\omega\right\vert^2 + 2\ N_i \rho^2 \left\vert {M_i \over
m_{3/2}} \right\vert^2 > 3\left(1 + \sum_A^{ }\nu_A \kappa_A
\rho^2 \right)
+ 2\
{{\rm d} g \over {\rm d} \ln T}
(T=1)
\eqno (24)
$$
We assume that the value of the gauge coupling, $<S^{-1}>$
is fixed by a more thorough treatment of the string symmetries. For $n = 1,$
one has
$$
\omega = 1 + {4 \pi ^2 \over g^2 C_2 (Adj G)}
\eqno (25)
$$
which is $> 3$ if $g^2 \sim 1/2$
(its experimental value if $G$ has $k=1$) and $C_2 (Adj G) < 4\pi^2.$
For $n>1,$ one can show that $\sum_i^{ }\beta_i U^i <\hat \beta
\sum_i^{ } U^i$
(since $U^i >0$), where $\hat \beta$ stands for any simple group $\hat G
\supset G.$
Combining group factors to increase the weight $k$ does not
reduce the value of $\beta$ either.
Therefore the condition (24) is always fulfilled and the minimum
of the potential is for $T=1.$ This is in desagreement
with the value (numerically) found in Ref.[9].
The potential at $T=1$ is given by (22) with $g(T=1) = 0.$
In the approximation (23) it becomes
$$
\hat V = \left\vert m_{3/2}\right\vert^2
\left(\left\vert\omega\right\vert^2 - 3 - \rho^2\sum_A^{ }\nu_A
\kappa_A \right) +2\sum_i^{ } N_i \rho^2
\left\vert M_i \right\vert^2
\eqno (26)
$$
For $g^2 \sim 1/2,$ $\hat V \ge 0.$
Therefore $\hat V$ vanishes at its absolute minimum which is
then at the point $m_{3/2} = 0,$ i.e.,
$U^i = 0,$ $X^i \to \infty.$ At least formally, this represents the absence
of condensation and, a fortiori, of symmetry breaking.

The discrepancy with respect to Ref.[9] can be traced back to the
absence of the terms containing $H^{\prime}\left(\left\vert M_i\right\vert
^2\right)$
in (22). They are introduced by the gap equation (21), which
as noticed before is not taken into account in the approach
of Ref.[9]. The net result in (24) is to flip the sign of the gaugino term
(in the approximation (23)). This difference explains the minimum
with $T\gg 1$ which was found in [9] for a reasonable choice of the cut-off
$\Lambda$ and $g^2 \simeq .4.$

In spite of that, it is worth discussing further the consistency of the
NJL approach to gaugino condensation. For this purpose we will assume
relatively large $g^2,$ so that $\omega^2 \simeq 1.$ In this case
(24) can be violated leading to a minimum at $T > 1.$ Let us assume
this to be the case and define
$$
\alpha = 3 \left( 1 - \left\vert g^2 (T)\right\vert ^2\right)
\eqno (27)
$$
at the minimum of the potential with respect to $T.$ Then,
the minimization with respect to the auxiliary fields
$U_i,$ with the constraints (15),
%
provides relations for the
auxiliary components in the form
$$ F^i_X+\beta_ i{F^i_U \over U_i}+{F_S \over 4}=0\ \ \ \ \ \ \ \ \ \ \ \ \ \
(i=1,...,n) \eqno (28) $$

It remains to minimize the potential with respect to the
$X_i$' s. It gives
(for $ U_i\not= 0),
$
$$ {F^i_X \over X_i+X^\ast_ i}+{F^i_U \over U_i}={F_S \over
S+S^\ast} -m^\ast_{ 3/2} \left[\alpha + \sum^{ }_ A\nu_ A\kappa_AH^{\prime}
\left(\kappa_A \left\vert m^2_{3/2} \right\vert \right) \right]\ \ \ \ \ \ \ \
\ \
(i=1,...,n) \eqno (29) $$

Then we eliminate $ F_S $ and $ F^i_U $ through (19) and (16) and work in terms
of
the
variables
$$ x_i={X_i+X^\ast_ i \over 2\beta_ i}\ \ \ \ \ \ \ \ \ \ M_i={F^i_X \over
2\beta_ ix_i} \eqno (30) $$

The scalar potential at the extrema can be written as,
$$ \eqalignno{ V & =2 \sum^{ }_ iN_i \left(2 \left\vert M_i \right\vert^
2H^{\prime} \left( \left\vert M^2_i \right\vert ,\Lambda \right)-H \left(
\left\vert M^2_i \right\vert ,\Lambda \right) \right) &  \cr  &  + \sum^{ }_
A\nu_ A \left(\kappa_A \left\vert m_{3/2} \right\vert^ 2\ H^{\prime}
\left(\kappa_A
\left\vert m_{3/2} \right\vert^ 2,\Lambda \right)-H \left(\kappa_A \left\vert
m_{3/2} \right\vert^ 2,\Lambda \right) \right) &  \cr  &  -2
\sum^{ }_ i\beta_
iM_i\hat U_i \left(1-x_i \right) & (31) \cr} $$
where
$$  M_i  ={-\beta_ ix_i\hat U_i^\ast \over N_iH^{\prime} \left(M^2_i
\right)}\ \ \ \ \ \    \ \hat U_i  ={M^3_c {\rm e}^{-1-{S \over 4\beta_ i}-{X_i
\over \beta_i}} \over
\left(S+S^\ast \right)^{1/2} \left(T+T^\ast \right)^{3/2}} \eqno (32)  $$
and the extremum conditions are:
$$ \eqalignno{ 2M_i \left(x_i-1 \right) & = \sum^{ }_ j\beta_ j\hat U_j^\ast
 \left[
\left(1+{S+S^\ast \over 4\beta_ j} \right) \left(1+{S+S^\ast \over 4\beta_ i}
\right) \right]+(\alpha +\varepsilon) m_{3/2}^\ast &  \cr
 m_{3/2} & =- \sum^{ }_ i\beta_ i\hat U_i
\ \ \ \ \ \ \varepsilon   = \sum^{
}_ A\nu_ A\kappa_AH^{\prime} \left(\kappa_A \left\vert m^2_{3/2} \right\vert
\right) &
  (33) \cr} $$

Let us first analyse the simplest  case with a simple group $ G, $ i.e., $
n=1. $ In
this case the above relations reduce to the simpler expressions,

$$ \eqalignno{ m_{3/2} & =-\beta\hat U &  \cr  &  ={-\beta M^3_c {\rm
e}^{-1-{S+S^\ast \over 8\beta} -x} \over \left(S+S^\ast \right)^{1/2}
\left(T+T^\ast \right)^{3/2}} &  \cr{ 2 \over N_G\rho^ 2}x(1-x) & =
\omega^2
-(\alpha +\varepsilon) = -2\delta &  \cr
M_{ 1/2} & =x{m_{3/2} \over N_G\rho^ 2} &  \cr V_{ {\rm min}} &
={2xm^2_{3/2} \over N_G\rho^ 2}M^2_{ {\rm Planck}} & (34) \cr} $$

where we have defined a parameter $\delta$ expected to be of ${\cal O}(1).$
The solution of (34) is
$$
x = {1-\sqrt{1 + 4\delta N \rho^2} \over 2} \simeq -\delta N \rho^2
\eqno (35)
$$
where the approximation is valid for
$N\rho^2 = (N\Lambda^2)/(32 \pi^2 M^2_{pl}) \ll 1.$ The condensate scale,
associated to chiral symmetry breaking is
$$
U = M_c^3 {\rm exp}\left( -{1 \over 4 \beta g^2} -1 + \delta N
\rho^2 \right)
\eqno (36)
$$
Supersymmetry breaking is characterized by the gravitino
mass
$$
\eqalignno{ \left\vert m_{3/2} \right\vert & = {\beta g^4 M_{pl}
\over 4 <\ T\ >^3} {\rm exp} \left(
 -{1 \over 4 \beta g^2} -1 + \delta N
\rho^2 \right) &  \cr
M_{1/2} & =\delta \ m_{3/2} & (37)\cr}
$$

{}From the formal point of view these results look consistent. But
the assumption $4\beta g^2 >1$ which is required to avoid the
constraint from (27) corresponds to
values of
$m_{3/2}$ too near to $M_{pl}.$ Instead, the physical value of $g^2,$
which if replaced in  (37) could allow for a better hierarchy between
$m_{3/2}$ and $M_{pl},$ is inconsistent with gaugino condensation
in the NJL approach.

Before we proceed to discuss the case $ n>1, $ let us compare the results
obtained here with the original calculation of gaugino condensation in the
framework of global supersymmetry in [3].  These authors write an
effective
superpotential for the composite field $ U $ in the form,
$$ W_{V.Y.}= \left({1 \over 4g^2}+\beta \ {\rm ln}{U \over \mu^ 3} \right)U
\eqno (38) $$
The first term replaces the gauge multiplet kinetic term. The second term
reproduces the anomaly in the conformal supercurrent, that comprehends
dilation,
superconformal and chiral symmetry currents, which would be generated by the
gaugino loops. The superpotential (38) is combined with some kinetic term
defined by a Kahler potential $ K_V \left(U,U^{\dagger} \right) $ to determine
an effective theory at
energies below the confinement scale. The gauginos are confined  and assumed
to be faithfully replaced at low energies by the $ U $-condensate.
Minimization
ot the scalar potential,
$$ \eqalignno{ V & = \left\vert F^U \right\vert^ 2 &  \cr F^U & =
\beta \ {\rm ln}{U \over\langle U\rangle}
\left[{\partial \over \partial U}{\partial \over \partial U^\ast} K
\left(U,U^\ast \right) \right]^{-1/2}
&  \cr\langle U\rangle &  =\mu^ 3 {\rm e}^{-{1 \over 2\beta g^2}-1} & (39)
\cr} $$
so that $ V=0 $ at the minimum $ U=\langle U\rangle , $ where supersymmetry
remains unbroken while
chiral symmetry is spontaneously broken. The scale $ \mu $ is the cut-off of
the
effective theory, which should be related to the confinement energy. The
results of [3]  are reproduced in the NJL approach if one assumes $ \mu
\simeq M_c $ and
set $ x=0, $ so that there is no loop-corrections to the scalar potential. In
order to allow for supersymmetry breaking one has to replace the gauge
coupling by the dynamical superfield $ S $ in $ W_{V.Y.}. $

This effective theory formulation of gaugino condensation is formally
reproduced in our approach by integration on the auxiliary superfields $ X_i $
in
(5) and then withdrawing the gaugino fields. This reproduces the
superpotential $ W_{V.Y.}. $

One can also consider the globally supersymmetric version of the NJL approach
considered here. The results are similar to (28) with $ F^S= \left(S+S^\ast
\right)^2U $ and  the
scalar potential,
$$ \eqalignno{ V & ={ \left\vert F^S \right\vert^ 2 \over \left(S+S^\ast
\right)^2} \left(1+\rho^ 2 \right)+2N_G\rho^ 2M^2 &  \cr x(1-x) & ={N\rho^ 2
\left(S+S^\ast \right)^2 \over \beta^ 2} &  \cr M & ={m_S \over 1-x}{
\left(S+S^\ast \right) \over 2\beta} &  (40) \cr} $$
where $ m_S=2F^S/ \left(S+S^\ast \right) $ is the dilatino mass. Again, with $
g^{-2} $ replaced back for $ S, $
the  minimum of $ V $ is for $ M=0 $ and no supersymmetry breaking.

For $ n>1 $ one has to solve  the set of equations (32) in the approximation
(23) and with $S + S^{\ast} < 4 \beta_i,$ namely
$$
 \eqalignno{ x_i \left(x_i-1 \right) & ={N_i\rho^2 \over 2 \beta_ iU_i}
- \delta \sum^{ }_ j\beta_ jU_j
&  \cr U_i & =M^3_c \ {\rm e}^{-{1 \over 2}-{S
\over 4\beta_i}-{X_i \over \beta_i}}   & \cr   M_i &=-{x_i\beta_ iU_i^\ast
\over N_i\rho^ 2} & (41) \cr} $$
where, by extension of (33), we note $2\delta = 1-\alpha-\epsilon.$

The expression for the potential $V_0 + V_1,$  before extremalization with
respect to
the $X_i$' s is:
$$
V \propto 2\delta \ \left\vert{ \sum_i^{ }\beta_i{\rm exp}
\left(-{ X_i \over \beta_i}\right)} \right\vert^2
+\sum_i^{ }{1 \over 2N_i\rho^2}\left(X_i + X_i^{\ast}\right)^2
\left\vert { {\rm exp}\left(-{X_i \over \beta_i}\right)} \right\vert^2
\eqno (42)
$$
In the case $2\beta_ig^2_i > 1,$ the first term
in (42) is negative, so that the minima are obtained for real values
of the $X_i.$ Though the explicit solutions have a involved dependence
on the $\beta_i$ they are basically analoguous to the $n = 1$
condensate. it is worth noticing that the potential (42) depends
on the phases of the $X_i$' s through its first term but they vanish
at the extrema.
A similar expression holds in the physically interesting case
$2\beta_ig^2_i < 1,$ but then all the terms in (42) are positive
as already remarked and the minimum is obtained in the
$X_i \to \infty$ limit.

As a final remark, the fermion condensation in the original NJL
model is due to an attractive four-fermion interaction. Notice that,
in formula (22), obtained after elimination of the auxiliary fields,
the tree-level scalar potential is $
\left(\alpha-\left\vert\omega\right
\vert^2\right)
\left\vert m_{3/2}\right\vert^2\ \propto \left(
\sum_i\ \beta_i \ U_i\right)^2,$ see relations (11) and (17), where $U_i$
stands for the gaugino
bilinears. Therefore, the effective four-fermion interaction after elimination
of constraints, appears to be  attractive (resp. repulsive) if $
\alpha\ > \left\vert\omega\right
\vert^2$
$\left({\rm resp.}\ \alpha\ <\ \left\vert\omega\right
\vert^2\right).$ This shows the essential r\^ ole of the negative contribution
specific to the supergravity potential, which is the only source
of attractive four-fermion interactions\footnote{$ ^4 $}{This remark is due to
E. Dudas. We thank him for useful discussions.}.
\vfill\eject
\centerline{{\bf REFERENCES}}
\vskip 24pt
\item{$\lbrack$1$\rbrack$} H.P. Nilles, {\it Phys. Lett.} {\bf B115} (1982) 193
, {\it Nucl. Phys.} {\bf B218}(1983) 366; S. Ferrara, L. Girardello and H. P.
Nilles, {\it Phys. Lett.} {\bf B125}(1983) 457.
\item{$\lbrack$2$\rbrack$}
J.-P. Derendinger, L.E. Iba\~ nez and H. P. Nilles, {\it Phys. Lett.} {\bf
B155}(1985) 65,
 {\it Nucl. Phys.} {\bf B267} (1986) 365;
M. Dine, R. Rohm, N. Seiberg and E. Witten, {\it Phys;
Lett.} {\bf B156}
(1985) 55.
\item{$\lbrack$3$\rbrack$} G. Veneziano and S. Yankielowicz, {\it Phys. Lett.}
{\bf
B113} (1982) 231.
For a review, see: D. Amati, K. Konishi, Y. Meurice, G.C. Rossi and G.
Veneziano, {\it Phys. Rep.} {\bf 162}(1988) 169, and references therein.
\item{$\lbrack$4$\rbrack$} J. Affleck, M. Dine and N. Seiberg, {\it Nucl. Phys.
} {\bf B241} (1984) 493;
 {\bf B256} (1985) 557;
T.R. Taylor, {\it Phys. Lett.} {\bf B164}
(1985) 43.

\item{$\lbrack$5$\rbrack$} S. Ferrara, N. Magnoli, T.R. Taylor and G.
Veneziano,{\it Phys. Lett.} {\bf B245}(1990) 409;
A. Font, L.E. Iba\~nez, D. L\" ust and F. Quevedo, {\it Phys. Lett.} {\bf
B245}(1990)401; P. Bin\' etruy and M. K. Gaillard,{\it Phys. Lett.} {\bf
B253}(1991) 119.

\item{$\lbrack$6$\rbrack$} N.V. Krasnikov, {\it Phys. Lett.} {\bf B193} (1987)
37;
J.A. Casas, Z. Lalak, C. Mun\~oz, and G.G. Ross, {\it Nucl. Phys.} {\bf B347}
(1990)243.
\item{$\lbrack$7$\rbrack$} T.R. Taylor, {\it Phys. Lett.} {\bf B252} (1990) 59.

\item{$\lbrack$8$\rbrack$} Y. Nambu and G. Jona-Lasinio, {\it  Phys. Rev.} {\bf
122}
(1961) 231.

\item{$\lbrack$9$\rbrack$} A. de la Macorra and G.G. Ross, {\it Nucl. Phys.}
{\bf
B404} (1993) 321.

\item{$\lbrack$10$\rbrack$} W. Buchm\"uller and S.T. Love, {\it Nucl. Phys.}
{\bf
B204} (1982) 213.
\item{\nobreak\ \nobreak\ \nobreak\ \nobreak\ } W. Buchm\"uller and U.
Ellwanger, {\it Nucl. Phys.} {\bf B245} (1984) 237.

\item{$\lbrack$12$\rbrack$} C.P. Burgess, J.-P. Derendinger, F. Quevedo,
and M. Quir\' os, Preprint CERN-TH/95-7; P. Bin\' etruy and M.-K. Gaillard,
private communication.

\item{$\lbrack$13$\rbrack$} E. Cremmer, S. Ferrara, L. Girardello and A. van
Proeyen, {\it Nucl.
Phys.} {\bf B212} (1983) 413.

\item{$\lbrack$14$\rbrack$} S. Ferrara, D. Lust, A. Shapere and S. Theisen,
 {\it Phys. Lett.} {\bf B225} (1989) 363.

\item{$\lbrack$15$\rbrack$} S. Coleman and E. Weinberg, {\it Phys. Rev.} {\bf D
7}
(1973) 1883.

\item{$\lbrack$16$\rbrack$} R. Barbieri and S. Cecotti, {\it Z. Phys.} {\bf C}
{\bf
17} (1983) 183.

\end